\documentclass[aps,  groupedaddress,  preprint,  showpacs, showkeys, floatfix]{revtex4}

\usepackage{graphicx}
\usepackage{amsmath}

\begin{document}



\title{Long-term drift laser frequency stabilization using purely optical reference.}

\author{A. Rossi}
\affiliation{INFM UdR Siena - Department of Physics, University of
Siena, Via Banchi di Sotto 55, 53100 Siena, Italy.}

\author{V. Biancalana}
\affiliation{INFM UdR Siena - Department of Physics, University of
Siena, Via Banchi di Sotto 55, 53100 Siena, Italy.}

\author{B. Mai}
\affiliation{INFN Sez. Ferrara - Department of Physics, University
of Ferrara, Via Paradiso 12, Ferrara, Italy.}

\author{L. Tomassetti}
\affiliation{INFN Sez. Ferrara - Department of Physics, University
of Ferrara, Via Paradiso 12, Ferrara, Italy.}

\date{July 2002}

\begin{abstract}

We describe an apparatus for the stabilization of laser
frequencies that prevents long term frequency drifts. A
Fabry-Perot interferometer is thermostated by referencing it to a
stabilized He-Ne laser (master), and its length is scanned over
more than one free spectral range allowing the analysis of one or
more lines generated by other (slave) lasers. A digital
acquisition system makes the detection of the position of all the
laser peaks possible, thus producing both  feedback of both the
thermostat and the error signal used for stabilizing the slave
lasers. This technique also allows for easy, referenced scanning
of the slave laser frequencies over range of several hundred MHz,
with a precision of the order of a few MHz. This kind of
stabilization system is particularly useful when no atomic or
molecular reference lines are available, as in the case of rare or
short lived radioactive species.



\end{abstract}

\maketitle

RSI VOLUME 73, NUMBER 7 JULY 2002, pages 2544-2548

\section{Introduction}

Many techniques have been developed to stabilize both the long
term and short term drift of laser frequencies. They mainly use
reference frequencies obtained by cavity resonances or atomic or
molecular transitions \cite{Demtroeder82}. Many commercial lasers
have built-in stabilization systems and analog or digital inputs
that can be used with custom, external referencing systems. We
developed a stabilization system to be used in experiments where
no atomic line references are available. In particular, it has
been designed to stabilize single-mode ring dye lasers and
Ti:sapphire lasers. The system provides a voltage which can also
be used as a feedback for other kinds of externally controllable
lasers, and in particular for diode lasers.

This problem has already been faced and successfully solved in the
framework of laser-cooling experiments involving short-lived
radioactive species \cite{Zhao98, jila94}, for which no atomic
vapor can be used. Similar applications may arise when rare
isotopes have to be excited. The peculiarities of our
stabilization technique also allow for easy scanning of the
stabilized frequency over broad ranges.

The technique makes use of an optical cavity whose length is
scanned over more than a free spectral range (FSR) by means of a
piezo actuator. The master laser and slave laser(s) beams are
collimated and simultaneously analyzed, so that a multiple peak
spectrum is observed. Finally, a reference signal is produced by
reading the relative positions of the observed peaks.

Our implementation simplifies the one described in \cite{Zhao98}
by using a thermal control of the optical cavity length, instead
of compensating the thermal drift with piezo actuators, thus
making   the use of high-voltage offset on the piezo unnecessary.
This choice is similar to the one reported in \cite{jila94} and it
makes the piezo response more constant in time. Actually, the
response of piezo actuators is non-linear, and the large values of
DC offset needed to compensate thermal drift of the cavity length
may dramatically change  the slope of the response and hence the
effect of the AC scanning signal. For this reason, differing from
\cite{Zhao98}, we do not need a continuous re-calibration of the
scan following the variation of the piezo response.

The same result could be achieved by using two separate actuators
for thermal compensation and for scanning, nevertheless our
solution is easier and also allows for simpler construction of the
cavity, because no fused quartz, invar or other materials with low
thermal coefficient are needed. In fact, the compensation of the
slow thermal drift of the cavity length does not need the fast
response of a piezo actuator to be accomplished, moreover the
thermal control does not suffer of the small range compensation
which is intrinsic in the piezo. In our case the optical cavity
was home-made in aluminium, and this choice also makes the cavity
alignment fast and cheap, with the use of a simple device produced
by a common tool machine.

The system is fully controlled by a computer program which
operates a commercial ADC-DAC card. The  program was developed in
order to achieve relatively fast operation, continuous control of
the cavity response, flexible adjustment of the feedback
parameters, and on-time visual monitoring of the laser spectra.

Analysis of the error signal has also been implemented in order to
characterize the performance of the system. All the digital
controls were developed in LabView, making use of either a 16 bit
or a 12 bit National Instruments card. No external electronics are
needed apart from a very simple voltage-to-current converter used
to supply the cavity heater. We used the program to stabilize only
one laser with respect to the He-Ne, but other laser lines can be
added and referenced with straightforward extensions of the
program. In particular with this system we plan to stabilize a
Ti:Sa ring laser working at $718~$nm and a diode laser working at
$817~$nm, which will be used as cooling and repumping lasers in an
experiment of magneto optical trapping (MOT) of Francium
\cite{atutovmoi01}.

\section{Experimental apparatus}
\begin{figure}
  \includegraphics[width=10cm]{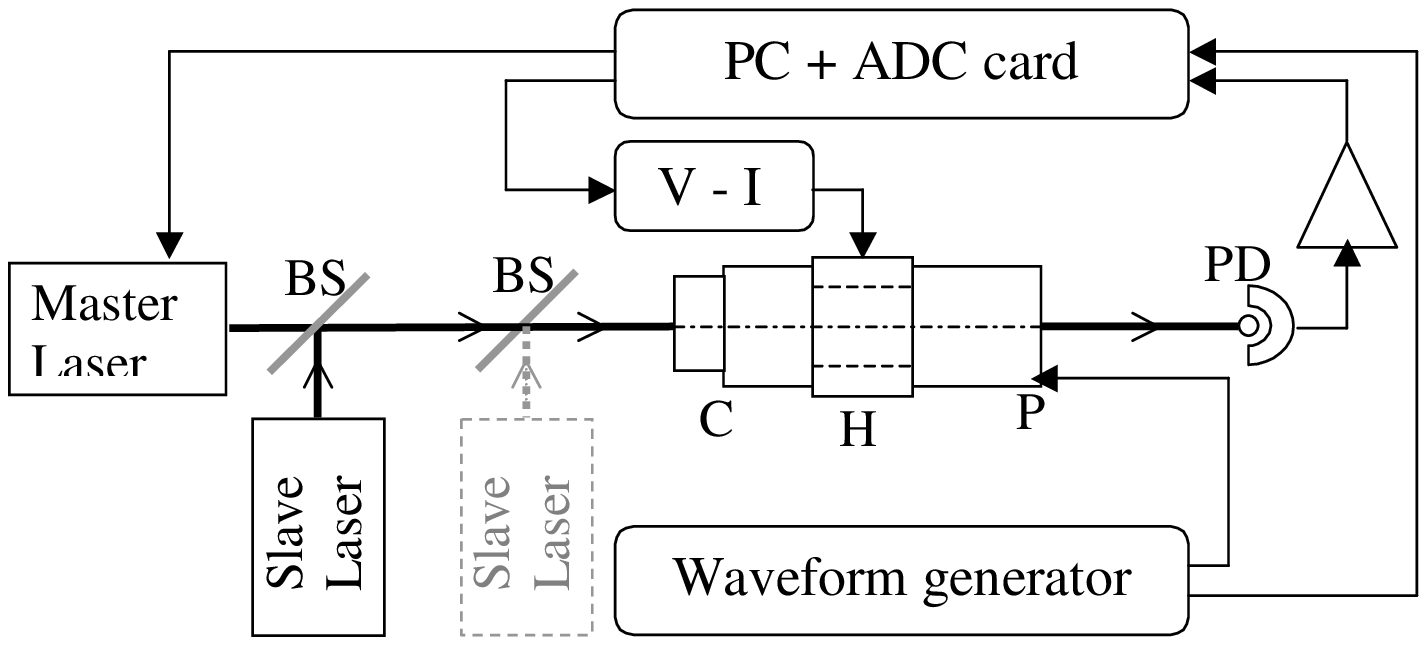}
  \caption{\label{fig:appsp} Schematic of the Apparatus.
  The beams of the master and slave lasers are made collinear, at
  the beam splitter BS. Similarly the beams of
  other slave lasers can be added.
  V-I is a voltage-to-current converter. The length of the cavity C is slowly adjusted
  by controlling the mount temperature through the heater H, and is dithered
  with a piezo P over a range of the order of half a wavelength.}
\end{figure}

The experimental apparatus is sketched in Fig.~\ref{fig:appsp}; it
consists of a Fabry-Perot (FP) confocal optical cavity
\cite{Demtroeder82, Born-Wolf}, on which all the lasers are
analyzed. An electric heater H (maximum power $10~$W) fed back by
the computer keeps the cavity at a temperature about 15$^\circ$ C
above the room temperature, stabilizing the cavity length to the
He-Ne line. The cavity length is then dithered over a range just
wider than one FSR by means of a piezo actuator (P) which is
directly driven by a $24~$V waveform generator. The transmitted
light is detected by a single amplified photodiode whose signal is
directly acquired by PC. The computer feeds back both the heater
through a voltage controlled current generator which thermally
stabilizes the cavity length, and the slave laser(s) with a
suitable error signal.

\subsection{The Fabry-Perot cavity}
\begin{figure}
  \includegraphics[width=12cm]{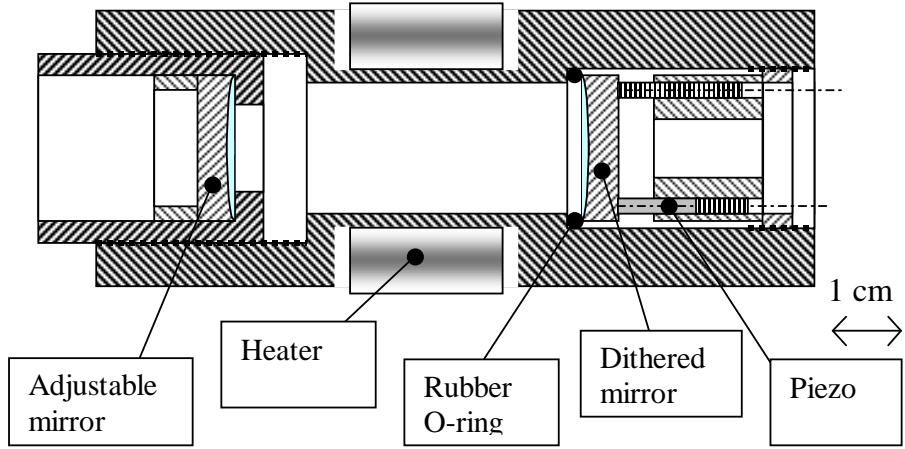}
  \caption{\label{fig:cavity} Schematic of the mechanical mount of the Aluminium cavity.}
\end{figure}

We built up several confocal Fabry Perot cavities having a
$1.5~$GHz free spectral range. The mirrors have a high
reflectivity for the $633~$nm He-Ne line and for other wavelengths
which are $780~$nm (diode laser for Rb lines), $590~$nm (dye laser
for Na lines), $718~$nm (Ti:sapphire for Fr lines). Each set of
mirrors has high reflectivity for two or more wavelengths. In this
paper we report results obtained with the $590~$nm device, by
which we stabilized a (Coherent) ring dye laser used for
experiments on sodium MOT's. The mirrors (1 inch, 5cm curvature
radius) were provided by CVI (TLM1 series). The measured finesse
is 70 for both the $633~$nm and the $590~$nm, consistent with the
declared reflectivity ($> 99\%$). The mirrors are mounted on an Al
tube adjustable in length, as represented in
Fig.~\ref{fig:cavity}.

Such a mount is home-made by a standard tool machine, allowing for
easy and effective centering of the optics at a very low cost. A
 thread is included in order to adjust the position of one mirror and to
 get an easy, precise match of the cavity length to operate in confocal regime. The
non-adjustable mirror is mounted on a rubber o-ring and it is
asymmetrically actuated by a low-voltage driven piezo (AE0203D08
Thorlab). The high stability of the confocal resonators makes this
peculiar approach possible, without relevant detrimental effects
on the quality of the observed spectra.

A part of the  aluminium tube -- where the heater is mounted --
has a thinner wall. This reduced thickness (about 1mm) is
essential to achieve a suitably fast response to the thermal
control. The heater consists of eight $2~$W, $100 ~ \Omega$
resistors, tightly fold onto the tube, powered with up to $10~$W.
The thermal contact between resistors and tube is improved by
means of thermal silicon grease for electronic purposes.

\subsection{Detection and acquisition}
A photodiode detects the transmitted light from the FP
interferometer and the photocurrent is converted into a voltage
signal {\it via} an operational amplifier, whose output is
directly acquired by computer.

In our present application, only one laser has to be stabilized
and a scan over a FSR is sufficient for the application, so that a
single photodiode is used and the peaks of both the master and the
slave lasers are acquired in a single trace. If more slave lasers
have to be stabilized, or if tunability on a wider range than a
FSR is requested, the different lines can  be sent on different
filtered photodiodes, and the program can be modified in order to
acquire more traces separately.

The signal from the photodiode is acquired as an array of 1000
values (this value can be adjusted), and a second ADC channel
acquires the signal driving the piezo actuator in a similar array.

The signal of the waveform generator is a triangular wave at a
frequency of about 10~Hz, (this is much lower than the measured
mechanical resonance of the piezo-mirror system, which is around
800~Hz). The triangular wave amplitude is 24~V peak-to-peak, and
it is enough to scan over more than a FSR.

No high-voltage signals are needed, and a commercial signal
generator is used. It is worth stressing that possible small
deviations in the signal slope or amplitude do not affect
precision as the transmittance is detected together with the
signal itself, so that the spectrum is plotted as a function of
the actual voltage applied to the piezo. This choice makes the
stability of the waveform generator  not crucial, nevertheless in
our case the generator is stable enough, and this detail turned
out to be not essential for the final performance.

The stabilization system provides a current which -- through the
heater -- keeps  the cavity length constant, and a voltage which
drives the slave laser frequency. These two signals are produced
according to different philosophies. Namely, the cavity length is
kept constant by keeping a transmission peak of the master laser
{\it close} to the  position selected by the operator, while the
slave laser peak is {\it precisely} kept at a given distance from
the master laser peak. In fact, the stabilization of the cavity
length does not demand absolute precision, having only the aim of
keeping the peaks within the scanned range, while the distance of
the peaks gives an error signal which is neither affected by
possible low-frequency noise in the triangular wave, nor by slight
oscillations of the cavity temperature (and hence average cavity
length).

As the cavity length is scanned over more than one master-laser
half-wavelength, it is possible to monitor two master-laser peaks.
The measured distance between two adjacent master-laser peaks
(FSR) allows for a precise calibration of the frequency axis on
the monitored spectra.

\subsection{Computer program and analysis of the performances}
\begin{figure}
  \includegraphics[width=8cm]{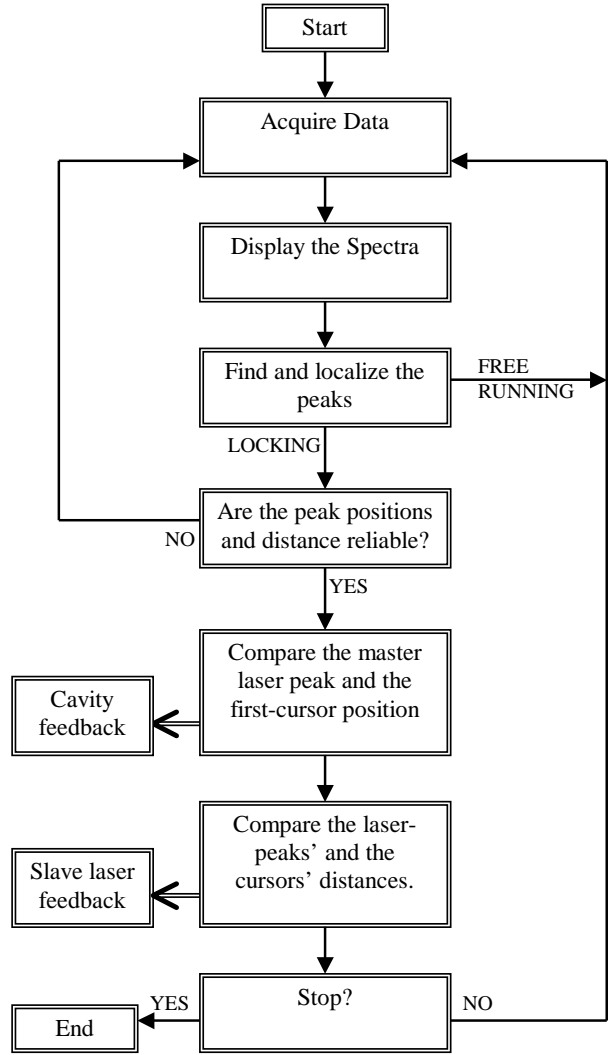}
  \caption{\label{fig:logic} The scheme summarizes the logic
  followed by the program in stabilizing the cavity length and the slave laser(s) frequency.
  For each scan of the piezo driving voltage, the spectra are displayed and analyzed.
  The peaks are detected and their position and relative distance is calculated.
  The feedback signal to the cavity thermal control is updated depending on the position
  of the master laser peak, while the feedback for the slave laser(s) is calculated
  using the distance of the master peak and slave peak(s). The cycle is directly restarted with no
  updating of the output when the detection of the peaks is not reliable, e.g. due
  to noise glitches, accidental shutting of a beam, etc.}
\end{figure}

The program was  developed using LabView and runs on a Pentium III
processor working at 1~GHz clock frequency. The computer works
with a 16-bit ADC card, model NI 6052E; we also checked the
program with a cheaper ADC card (namely the 12-bit NI LabPC+)
obtaining comparable performances. The logic of the program is
summarized in Fig.~\ref{fig:logic}.

The program acquires both the slope of the waveform generator and
the FP pattern, then calculates -- in terms of the voltage on the
slope -- the absolute position  of the He-Ne peak and the distance
between the He-Ne peak and the slave laser peak. By comparing
these two data with the nominal values, the program calculates the
two feedback signals for the FP thermal stabilization and the
slave laser stabilization. The nominal values are set by means of
two cursors. The position of the first cursor sets the position of
the master laser peak, and hence the cavity length, while the
second cursor sets the position of the slave laser(s) peak, and
hence the final frequency.

Each acquisition consists in n scans (we used $n=1000$) and the
program takes 10 acquisitions per second, while refreshing the
feedback signals at the same rate. The sampling rate is 20000
samples per second. All the data and the corrections to the
feedback signals are reported on the screen, which is also
refreshed after each acquisition.

If the program finds  problems in calculating the peaks' position
(e.g. due to glitches originated by the laboratory electric noise
affecting the photodiode signal), it leaves the feedback signals
unchanged and goes to the next acquisition; if the problem
persists for several cycles the program alerts the user, with a
permanent alarm which is manually resettable, and a counter
reports the number of cycles which gave problems.

The feedback signals consist of two voltages varying between -5
and 5~V. One of these signals is sent directly to the external
control of the slave laser, the other is converted by  a
voltage-to-current amplifier to supply the cavity heater. The
algorithms which calculate the correction of the two signals are
different: the thermal stabilization needs only a correction
proportional to the deviation of the He-Ne peak from the selected
position, and the thermal capacitance of the FP cavity provides an
integration of the discontinuities of the output voltage by
itself. The stabilization signal sent to the slave laser needs a
more advanced algorithm: the program  must integrate the output
signal to keep  the laser frequency stable over the short-time
range. This is achieved by limiting the maximum slope of the
correction signal, which can be optimized by the user, according
to the typical drift velocity encountered in the internal
stabilization system of the laser.

When necessary, the program makes it possible to scan the laser
voltage from -5~V to 5~V, searching for a reference signal (for
example the fluorescence of the MOT) to set  the unknown value of
the right laser voltage automatically; obviously the user has to
stop the laser stabilization before running the program with this
aim.

\begin{figure}
  \includegraphics[width=12cm]{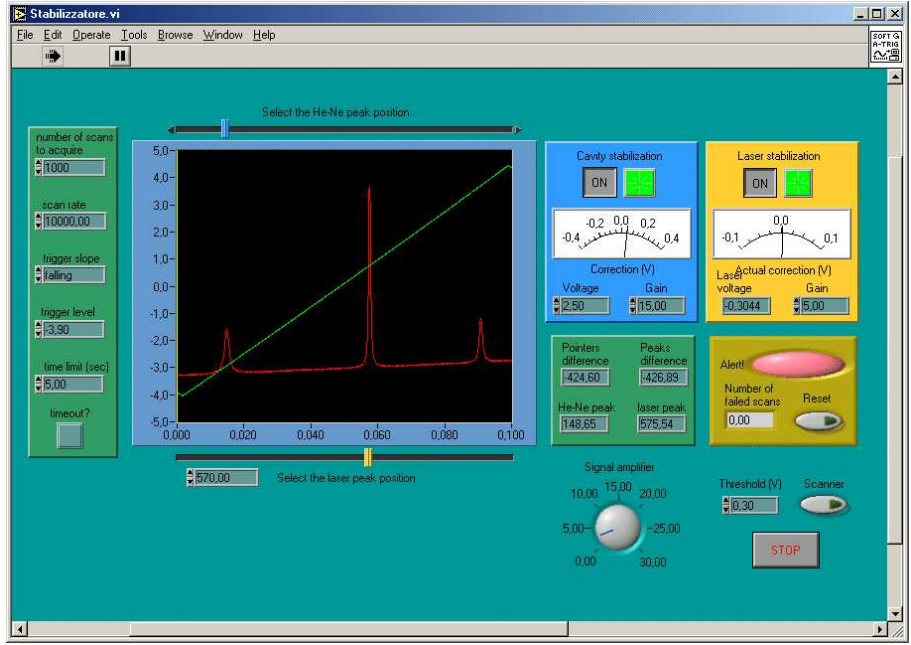}
  \caption{\label{fig:schermo} This is the window shown to the user, which allows
  him to control the program. On the left side there are controls for the scan rate, the number
  of scans (abscissa points), the trigger, and trigger time-out. The spectra are shown in the
  central panel together with the piezo ramp. Two cursors above and below this central panel
  allow for placing and displacing of the peak positions.  On the right
  side, on the upper part there are two panels with
  switches  for starting the stabilization of the cavity and slave laser respectively,
  and two indicators showing the actual error signals. In the lower part the nominal and actual peak positions
  are reported in one panel, while on the other panel an alert goes on when locking fails,
  and a counter reports the number of failed scans. Finally a knob makes it possible to adjust
  the gain on the photodiode signal, and the trigger threshold can be set numerically
  from an input close to the "stop" button. Another button starts the scan operation.}
\end{figure}

The scan operation will have an important role in the application
in the francium experiment. In fact, the locking procedure will
start by using a wave-meter in order to set the laser frequency in
resonance with an uncertainty of several hundred MHz. Then the
control of the laser will be passed to the program, which will
scan the frequency over a range just wider than the wave-meter
uncertainty, looking for the exact resonance. During the scan, the
frequency will be referenced with respect to the He-Ne peaks. Once
a known resonance is found, the program will provide an absolute
frequency scale, which will keep being available as long as the
He-Ne laser stabilization system of the cavity is kept on. Finally
the frequency will be locked to the exact value, and the program
will provide the error signal necessary to maintain  long-term
stability. A user-friendly graphic interface (see
Fig.~\ref{fig:schermo}) makes use of a few numerical controls and
cursors, which allow for
\begin{itemize}
\item locking of the cavity with respect to the He-Ne FP peaks;
\item scan of the slave frequency in a given interval;
\item fast setting and precise adjustment of the stabilization
point.
\end{itemize}

\section{Results}
\begin{figure}
  \includegraphics[width=10cm]{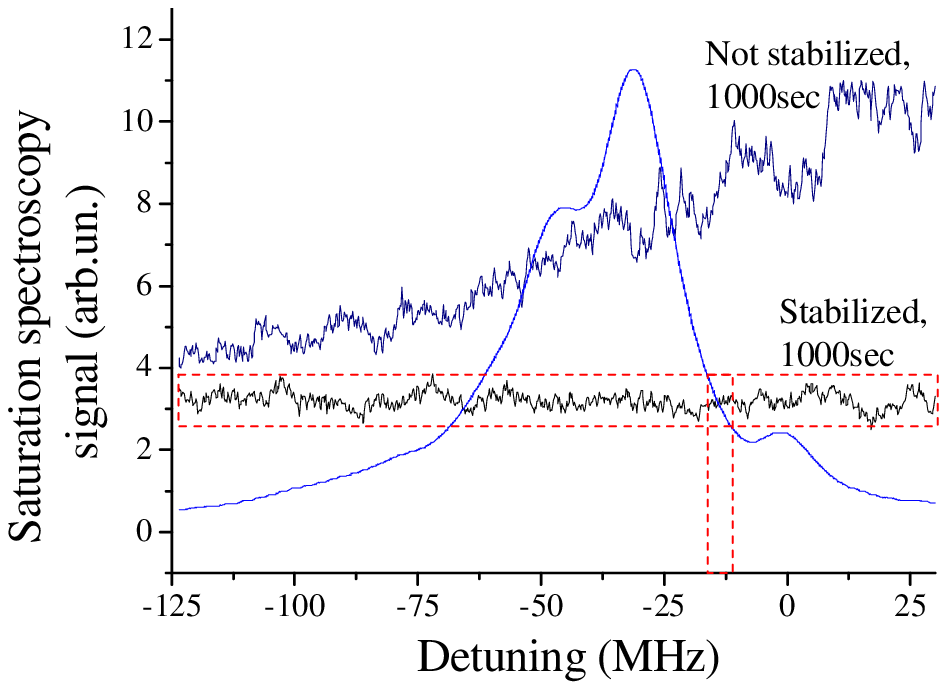}
  \caption{\label{fig:results} The achieved frequency stability of the laser is
  estimated by measuring the fluctuation of the saturation spectroscopy signal on a Na cell.
  The figure shows a spectrum and the location at which the laser was stabilized
   (red wing of the $D_2$, $F=2 \to F'=3$ transition).
  Also two traces are reported of $1000~$s obtained in conditions of
  stabilized and not-stabilized
  operation respectively. The boxes show that the frequency varies
  within a range of 4~MHz when the stabilization system is on.
  This is probably an overestimation due to uncontrolled fluctuations in the reference cell,
  in fact the fluorescence of the MOT was extremely stable, even using detuning
  close to the edge of the trapping range.}
\end{figure}

Fig.~\ref{fig:results} reports the saturation spectroscopy signal
obtained by scanning the laser frequency over the $D_2$, $F_g=2
\to F_e=1, \,  2, \, 3$  lines of sodium \cite{atutov01}. Two
traces of the same signal are also reported as obtained by keeping
the frequency at a nominal fixed value, either with the
stabilization system on or off. The recording time for these two
traces is $1000~$s.

Short term deviation of the laser frequency as deduced from these
plots is definitely less than $\pm 2~$MHz, as graphically shown by
the dashed boxes. Statistical analysis on the  trace recorded in
stabilized conditions shows that the ratio between the standard
deviation and the range of values is less than 0.15, corresponding
to a standard deviation of 600~kHz in the frequency scale.

These estimations of the frequency deviation may be larger than
the actual values, because other effects  produce some noise on
the saturated spectroscopy signal. In fact, a relevant noise level
in is also visible on the wing of the saturation spectroscopy
signal, which at detunings larger than 100~MHz, should be
nominally zero.

The short term (few seconds) deviations of the frequency appear
with similar features both in the stabilized and in the
not-stabilized operation, so that they are probably intrinsic in
the internal stabilization system of the laser; the comparison of
the two curves in Fig.~\ref{fig:results} show that they are only
partially reduced by our external device, which on the contrary
definitely fixes  the long-term drift.

We performed a cross-check of  stability using the fluorescence
signal of the atoms trapped in a MOT using the $F_g=2 \to F_e=3$
transition of the $D_2$ line of the sodium. By displacing the
slave laser frequency in steps of 2~MHz, it was possible to
evaluate in 14~MHz the total spectral width of the trap.

As also reported in \cite{atutov01}, at the high-frequency side
the trap abruptly disappears. In our observation we were able to
keep the trap unstable on that condition, with a fluorescence
signal significantly  lower than the maximum, and essentially
stable in time.

\section{Discussion and Conclusion}
A cheap and easy-to-use set-up was developed which, coupled with
an efficient computer program, a commercial ADC-DAC card, and a
standard confocal FP interferometer allows for active long term
stabilization of one or more laser frequencies. A long term
stability better than $4~$MHz was demonstrated, and a FWHM of
$1.2~$MHz was achieved in the statistical distribution of the
stabilized frequency. The short term stability keeps being given
by the internal, fast stabilization system of the ring dye laser
used as a slave.

The peculiar approach which stabilizes the relative peak position
of the master and slave laser(s) reduces the effect of low
frequency noise in the driving voltage applied to scan the piezo,
and makes   ultra high accuracy on the cavity length stabilization
unnecessary, so that only a slow and not high precision thermal
feedback must be used, in order to always keep  given interference
orders within the scanned range.

The final stability which is achieved with this technique is in
excess with respect to the demands of the experiment to which this
technique was applied. Further improvements could be made by using
a longer cavity and separate photodiodes for the acquisition of
the peaks corresponding to different lasers, and setting up FP
with a smaller FSR. In fact with the single photodiode operation
the adjustable range is limited by one FSR, as no superposition of
peaks is allowed.

We address to \cite{jila94} for a detailed analysis of the
limitations on  long term  stability which are set by the effects
of atmospheric pressure and humidity, whose variation may
introduce errors due to different changes of air refraction index
for the different wavelength of the laser used. Variations in
atmospheric temperature are also potentially critical, but this
problem is definitely overcome in setups  where a close cavity is
thermally stabilized. In our case the difference in the two
wavelengths is smaller than the one reported in \cite{jila94},
thus reducing the effect due to dispersive refraction index of
humidity.

\begin{acknowledgments}
The authors thank all the colleagues of the Siena, Ferrara and
Legnaro laboratories for the encouragements and the useful
discussions. Evro Corsi and Alessandro Pifferi are thanked as
well, for their effective technical support.
\end{acknowledgments}

\end{document}